\documentclass{book}    

\usepackage{piers}  
\usepackage[]{graphicx}
\graphicspath{{Figures/}}
\usepackage{graphicx}

\usepackage{amsmath}
\usepackage{amssymb}
\usepackage{epsfig}
\usepackage{cite}
\usepackage{color}
\usepackage{balance}
\usepackage{pgfplots}
\usepackage{wrapfig}
\usepackage{lipsum}
\usepackage{hyperref}
\usepackage{gensymb}

\begin{document}

\title{Predicting Solar Irradiance in Singapore}
\maketitle

\author      {F. M. Lastname}
\affiliation {University}
\address     {}
\city        {Boston}
\postalcode  {}
\country     {USA}
\phone       {345566}    
\fax         {233445}    
\email       {email@email.com}  
\misc        { }  
\nomakeauthor

\author      {F. M. Lastname}
\affiliation {University}
\address     {}
\city        {Boston}
\postalcode  {}
\country     {USA}
\phone       {345566}    
\fax         {233445}    
\email       {email@email.com}  
\misc        { }  
\nomakeauthor

\begin{authors}

{\bf T. A. Fathima}$^{1}$, {\bf Vasudevan Nedumpozhimana}$^{2}$, {\bf Yee Hui Lee}$^{3}$, \\{\bf Stefan Winkler}$^{4}$, {\bf and Soumyabrata Dev}$^{2,5,6*}$\\
\medskip

$^{1}$Indian Institute of Technology Bombay, Powai, Mumbai, India\\

$^{2}$ADAPT SFI Research Centre, Dublin, Ireland\\

$^{3}$Nanyang Technological University Singapore, Singapore 639798\\

$^{4}$ Department of Computer Science, National University of Singapore, Singapore 117417\\

$^{5}$School of Computer Science, University College Dublin, Ireland.

$^{6}$Beijing-Dublin International College, Beijing, China.

$^{*}$ Presenting author and corresponding author

\end{authors}

\begin{paper}

\begin{piersabstract}
Solar irradiance is the primary input for all solar energy generation systems. The amount of available solar radiation over time under the local weather conditions helps to decide the optimal location, technology and size of a solar energy project. We study the behaviour of incident solar irradiance on the earth's surface using weather sensors. In this paper, we propose a time-series based technique to forecast the solar irradiance values for shorter lead times of upto 15 minutes. Our experiments are conducted in the tropical region \emph{viz.} Singapore, which receives a large amount of solar irradiance throughout the year. We benchmark our method with two common forecasting techniques, namely persistence model and average model, and we obtain good prediction performance. We report a root mean square of $147$ W/m$^2$ for a lead time of $15$ minutes. 
\end{piersabstract}

\psection{Introduction}

The growing energy demand  due to the technological advancements and population explosion is one of the biggest challenges we are facing these days. If we continue to depend only on fossil fuels for our energy demand, there are greater chances for the depletion of fossil fuels. Also there is an increasing concerns on green house gas emissions due to the burning of fossil fuel like coal. It has been reported that greenhouse gases have greater impact on climate system by global warming~\cite{Darkwahetal2018}. In these circumstances, a reliable, renewable energy source like solar energy is considered as a promising source for managing the long-term issues in the energy crisis. Solar irradiance is the primary input for all solar energy generation systems. The amount of available solar radiation over time helps to decide the optimal location, technology and size of a solar energy project. Thus, the annual solar energy output of a photovoltaic system depends on the annual average solar radiation. A few cloudy, rainy days can have adverse effects on the efficiency of photovoltaic systems as the efficiency of solar energy is highly dependant on the weather conditions. The forecasting of solar irradiance in advance helps to estimate the annual average solar radiation and it also helps in the smooth operation of switching circuit of the grid connected photovoltaic cells without much voltage fluctuations. The present work is focused in the forecasting of solar irradiance using a univariate time series forecasting technique called Triple Exponential Smoothing (TES).

\psection{Triple Exponential Smoothing}
Exponential Smoothing is one of the widely used technique for forecasting univariate time series data. The forecast will be the weighted average of the past observations in which the weight will decay exponentially as the observations gets older~\cite{hyndman2018}. In other words, the associated weights will be higher for the most recent observations, and lower for historic observations. The TES is an extension of exponential smoothing that explicitly adds support for trend and seasonality to the univariate time series.

There are primarily three parameters that are considered for the forecast using TES method. The first parameter $\alpha$ controls the rate at which the influence of the observations at prior time steps decay exponentially. Large values of $\alpha$ mean that the model pays attention mainly to the most recent past observations, whereas smaller values mean more of the history is taken into account when making a prediction. The second parameter $\gamma$ is to control the decay of the influence of the change in trend of the time series. The third parameter $\beta$ controls the influence on the seasonal component of time series. The TES method~\footnote{\url{https://www.itl.nist.gov/div898/handbook/pmc/section4/pmc435.htm}} is described as follows:

\begin{eqnarray}
S_t  =  \alpha \frac{y_t}{I_{t-L}} + (1-\alpha)(S_{t-1}+b_{t-1})  \,\,\,\,\,   \mbox{Overall smoothing} \\
b_t  =  \gamma (S_t - S_{t-1}) + (1 - \gamma)b_{t-1} \,\,\,\,\,  \mbox{Trend smoothing} \\
I_t  =  \beta \frac{y_t}{S_t} + (1 - \beta) I_{t-L}  \,\,\,\,\, \mbox{Seasonal smoothing} \\
F_{t+m}  =   (S_t + m b_t) I_{t-L+m}  \,\,\,\,\, \mbox{Forecast},
\end{eqnarray}

where $t$ is the time index, $y$ is the measured solar irradiance value, $S$ is the smoothed record, $b$ denotes the trend factor, $I$ denotes the seasonal index, $F$ is the forecasted solar irradiance value with a lead time of $m$, and $L$ is the number of periods in a season. The Triple Exponential Smoothing technique has already been used in different time series data~\cite{dev2018solar, Papacharalampousetal2018, Rahmanetal2016, manandhar2019predicting}. In ~\cite{dev2018solar}, the TES method has been used for solar irradiance prediction for the region of Utrecht, the Netherlands. In this paper, we use this technique for forecasting the solar irradiance in the tropical region \emph{viz.} Singapore, particularly for shorter lead time (cf.\ Fig.~\ref{fig:short-time}).

\begin{figure}[htb]
\begin{center}
\includegraphics[width=0.7\textwidth]{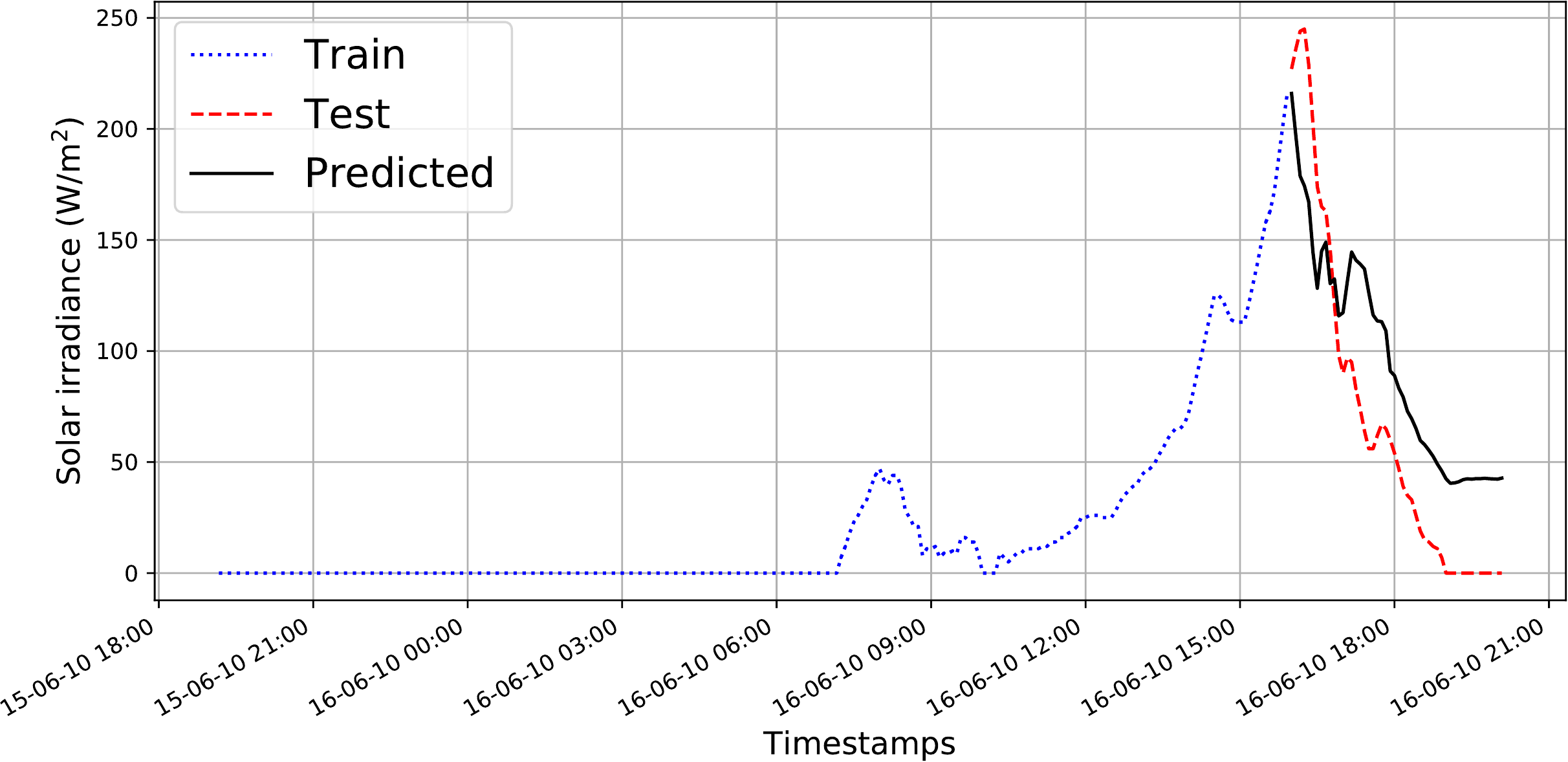}\\
\includegraphics[width=0.7\textwidth]{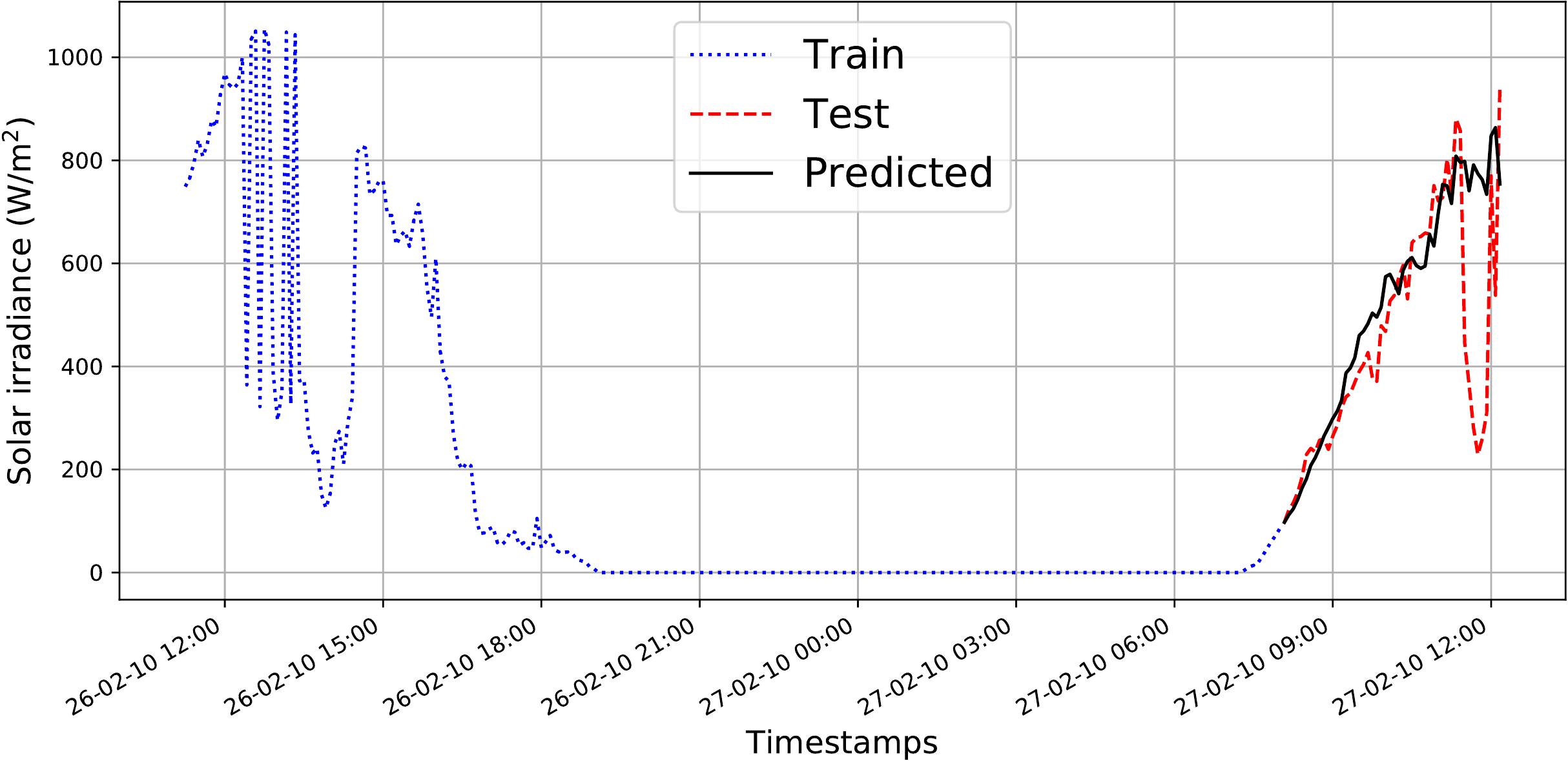}
\caption{Prediction of solar irradiance for shorter lead time. For the sake of brevity and clarity of figures, we only plot the previous $250$ data points from the training set.
\label{fig:short-time}}
\end{center}
\end{figure}

\psection{Results and Experiments}
In this section, we provide a detailed discussion on the forecasting of solar irradiance values using exponential smoothing.

\psubsection{Data}
We perform our experiments in Singapore. Our weather measurements are recorded at the rooftop of the university building at Nanyang Technological University Singapore, located at $1.3\degree$N, $103.68\degree$E. We use Davis Instruments 7440 Weather Vantage Pro I with a tipping rain gauge to record temperature, humidity, wind-speed, solar irradiance, dew point temperature and rainfall rate. All these measurements are recorded with a resolution of $5$ minutes.

\psubsection{Qualitative Evaluation}

In this section, we provide a time series analysis of the prediction of solar irradiance. We use a training set of $2000$ observations to train the TES model. We set the seasonal period as $288$ in the TES model. We perform a subjective evaluation of the proposed method for both short and long lead times. Figure~\ref{fig:short-time} illustrates the performance for shorter lead time, where we forecast the subsequent $50$ observations. We observe that the predicted solar irradiance can estimate the peaks and the gradual fall with a good degree of accuracy. Our predicted solar irradiance can also estimate the gradual rise of solar irradiance in the early hours of 27-Feb-2010 (cf.\ Fig.~\ref{fig:short-time}). 

Similarly, in Fig.~\ref{fig:long-time}, we illustrate the performance of our proposed method for longer lead time. We attempt to forecast the future $500$ observations. The several peaks and troughs of solar irradiance are properly captured by our method. The global trend is captured with a good degree of accuracy, but it fails to capture the intermediate rapid fluctuations. The general trend of solar irradiance for a day-ahead prediction, as depicted in 27-Feb-2010 and 28-Feb-2010 are captured properly by the proposed method (cf.\ Fig.~\ref{fig:long-time}). We observe that TES is good at forecasting similar patterns, but could not forecast the short-term fluctuations accurately. 

\begin{figure}[htb]
\begin{center}
\includegraphics[width=0.7\textwidth]{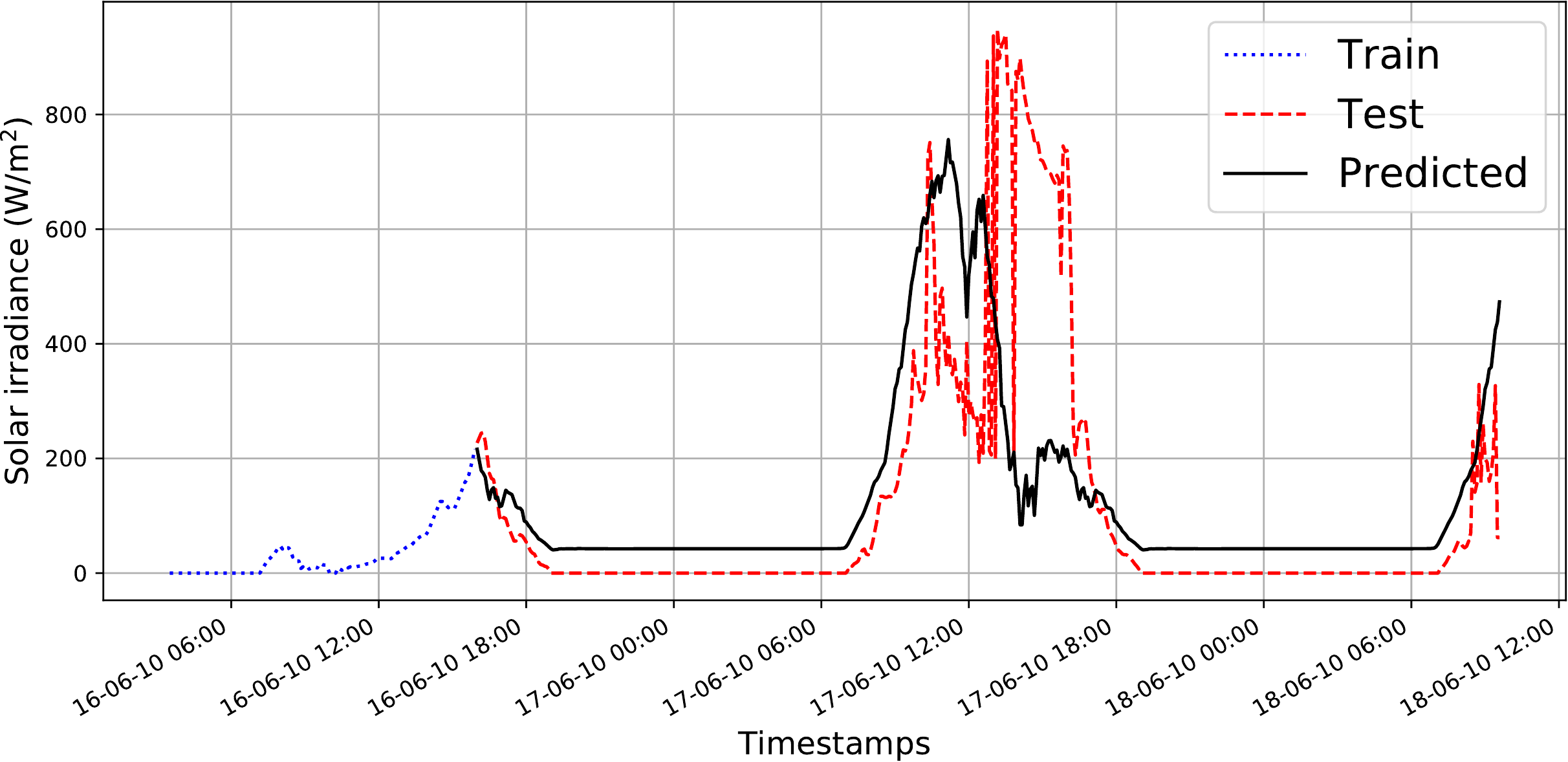}\\
\includegraphics[width=0.7\textwidth]{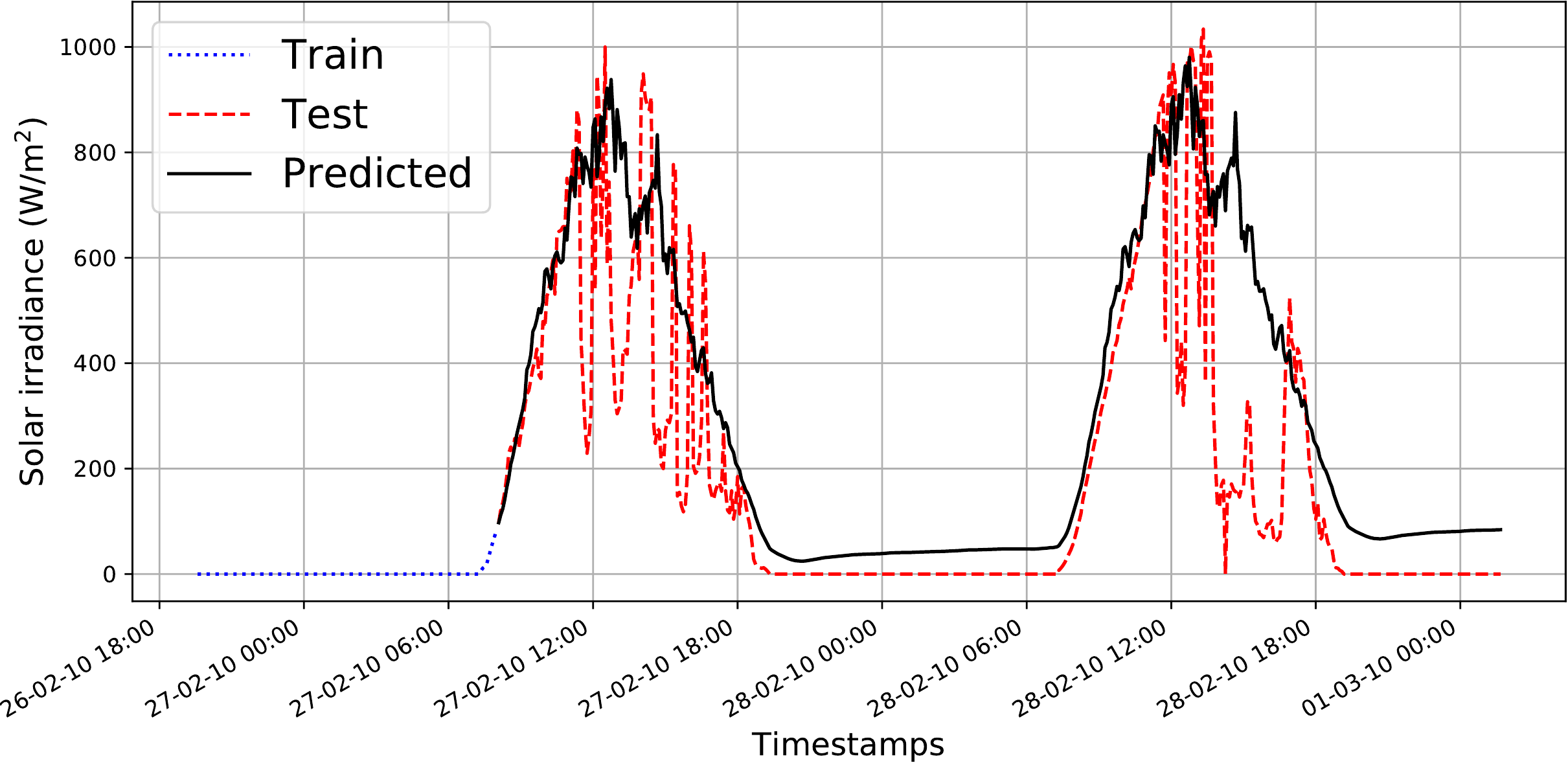}
\caption{Prediction of solar irradiance for longer lead time. For the sake of brevity and clarity of figures, we only plot the previous $150$ data points from the training set.
\label{fig:long-time}}
\end{center}
\end{figure}

\psubsection{Quantitative Evaluation}
In addition to the subjective evaluation of our proposed forecasting approach, we also provide an objective evaluation. We benchmark our method with two baseline approaches -- persistence model and average model. The persistence model assumes that the solar irradiance value at time $t+1$ is the same as recorded at time $t$. This model assumes that the atmospheric conditions remain same throughout the forecast period. The average model forecasts the future observations by averaging the solar irradiance measurements in the training period. We compute the Root Mean Square Error (RMSE) value between the actual solar irradiance and the predicted solar irradiance. The RMSE values are computed for lead times of $5$, $10$ and $15$ minutes. As solar irradiance is zero during the nighttime, we consider observations from $07:00$ am till $06:00$ pm. In order to remove any sampling bias, we perform $10$ distinct experiments from our dataset of weather records. Table~\ref{table:compare} summaries the comparative results of the different benchmarking algorithms. Amongst the different benchmarking algorithms, we observe that the average method performs the worst. This is because the average method computes the global average across all the observations in the training stage, and estimates an erroneous value of solar irradiance. The persistence method is better than average method, mostly for shorter lead times. Our proposed method using TES perform the best amongst all benchmarking methods, across all the lead times.

\begin{table}[htb]
\centering
\normalsize
\caption{RMSE (W/m$^2$) for varying lead times of the benchmarking algorithms. The reported values are the average obtained from $10$ experiments.}
\begin{tabular}{cccc}
\hline
Lead Time & Proposed & Persistence & Average \\
\hline 
5 min   &  29.28    & 55.30  & 298.63 \\
10 min & 111.71    & 167.89     & 200.06  \\
15 min  & 147.32    & 188.17       & 322.96 \\
\hline
\end{tabular}
\label{table:compare}
\end{table}

\psection{Conclusion and Future Work}
In this paper, we have proposed a time-series based technique for forecasting solar irradiance in Singapore. We use triple exponential smoothing technique to model the seasonality and variability of the solar irradiance fluctuations. The RMSE of the proposed method is lower for all lead times, as compared to the benchmarking algorithms. In the future, we intend to provide a detailed statistical analysis of the performance of the proposed forecasting technique by using weather records of larger time period. We also intend to use other meteorological sensors~\cite{manandhar2018systematic} and ground-based cameras~\cite{WAHRSIS} to further improve the forecasting accuracy. 

\ack
The ADAPT Centre  for  Digital  Content  Technology  is  funded  under  the  SFI Research Centres Programme (Grant 13/RC/2106) and is co-funded under the European Regional Development Fund.

\bibliographystyle{IEEEbib}

\end{paper}

\end{document}